\input amstex
\input vanilla.sty
\pagewidth{16 cm}
\pageheight{24 cm}
\TagsOnRight
\baselineskip 18.5pt
\define \la {\lambda}
\define \part {\partial}
\define \si {\sigma}

\centerline{{\bf
Coupled Integrable
Systems Associated with a Polynomial}}
\centerline{{\bf
Spectral Problem
and their Virasoro Symmetry Algebras}}

\vskip 5mm
\centerline{Wen-Xiu MA$^{\dagger \ddagger}$ and Zi-Xiang ZHOU$^{\dagger *}$}

\vskip 2mm 
\centerline  {$^{\dagger}$Institute of Mathematics, Fudan University, 
Shanghai 200433, 
P. R. of China}
\centerline{$^{\ddagger}$FB Mathematik-Informatik, Universit\"at-GH Paderborn,
 D-33098 Paderborn, Germany}
\centerline  {$^*$Department of Mathematics, 
University of Pennsylvania, Philadelphia, 
PA19104, USA}

\vskip 6mm
\centerline {Abstract}
\vskip 2mm
\smallpagebreak
{\narrower
{\it
An isospectral hierarchy of commutative integrable systems associated 
with a polynomial spectral problem is proposed. The resulting 
hierarchy possesses a
recursion structure controlled by a hereditary operator. The nonisospectral
flows generate the time first order dependent symmetries of 
the isospectral hierarchy, which 
 constitute Virasoro symmetry algebras
 together with commutative symmetries. }
\par}
\smallpagebreak

\vskip 5mm

\centerline {{\bf 1. Introduction}}

\vskip 3mm

The study of integrable systems has been approached from various
points of view. Mathematically, a very successful analytic method
is the inverse scattering transform (IST), which has become
the standard method for solving nonlinear systems.
This method is based mainly upon two linear spectral problems, i.e.
a Lax pair$^{1)}$, connected with nonlinear systems.
At the beginning of development of IST, it is required that the
eigenvalues of the associated spectral operator be left invariant
when the potential evolves according to the nonlinear systems.
Subsequent investigation (see, for example, Ref. $2)\,$)
has shown that the nonisospectral nonlinear
systems associated with Lax pairs are also important and 
may be yet solved  by IST.

With the help of Lax pairs, one has  
generated some integrable 
systems of nonlinear differential and difference equations
important in physics fields.
A good skeleton to do so is Sato theory$^{3)}$, in which 
commuting flows and $\tau $-functions are presented simultaneously.
However Sato theory is a general theory, though it involves 
many integrable systems
of soliton equations 
as reductions, and thus 
it is still important to find concrete specific integrable systems, especially
coupled integrable systems$^{4)}$. 
In addition, the problem of constructing integrable systems itself
is also interesting both from the mathematics as well as 
the physics point of view.

In the present paper, we would like to present a class of
isospectral 
($\la _{t_m}=0$)
 coupled  integrable systems
by introducing a special spectral problem with the polynomial
form of the spectral parameter $\la $. All of those systems
can be solved by following the standard IST.
A simple calculation allows us to characterize them
iteratively so that the whole 
integrable hierarchy may be written as a nice and 
neat form. Nonisospectral ($\la _{t_n}=\la ^{n+1}$) flows generate  
a centerless Virasoro master symmetry algebra for each 
system in the isospectral hierarchy. 
Therefore the resulting isospectral hierarchy provides us with some typical
integrable systems of soliton equations.
However, we haven't been able to find out Hamiltonian structures and to give
rise to usual Darboux transformations$^{5)}$  
for the proposed isospectral hierarchy.

\vskip 4mm
\centerline {{\bf 2. Isospectral integrable systems}}
\vskip 3mm

Let us 
start from the spectral problem 
$$\phi_x=U\phi=U(u,\lambda )\phi ,\ 
\phi=( \phi_1, \phi_2)^T,
\ u=(v_0,v_1,\cdots,v_{q-1})^T,\tag2.1a$$
with the spectral operator $U$ being of the polynomial form of $\la $
$$
 U=Q\si _1+i\si _2+\si _3=\left (\matrix 1
&Q+1\\ Q-1& -1 \endmatrix \right ), 
\ Q=Q(u,\la )= \sum _{i=0}^q
v_i\la ^i,\ v_q=-1.\tag2.1b$$
Here $q$ is an arbitrary integer and 
the $\si _j$ are $2\times 2$ Pauli matrices. 
This spectral problem is a generalization of one appearing in 6) and has a
multi-component potential $u$. Our discussion here will be focused on   
constructing integrable systems.

In order to derive the related hierarchy of isospectral ($\la _{t_m}=0$)
systems, we introduce the auxiliary problem associated with (2.1)
$$\align &\phi_t=V\phi=V(u,\lambda )\phi ,\tag2.2a
\\ \vspace {1mm} & V=a\si _1 +bi\si _2+c\si _3=
\left (\matrix c &a+b\\ a-b& -c \endmatrix \right ). \tag2.2b\endalign
$$
It is readily worked out that 
$$[U,V]=(2b-2c)\si _1+(2a-2Qc)i\si _2+(2a-2Qb)\si _3.$$
Therefore we see that the compatibility condition
$U_t-V_x+[U,V]=0$
of the Lax pair (2.1) and (2.2)
becomes 

\vskip 2mm
\line
{\hbox to 0pt {\hss}
\hss $\displaystyle
\left \{\matrix \format \l\\
Q_t-a_x+2b-2c=0,\\ 
\vspace {1mm}
-b_x+2a-2Qc=0,\\
\vspace {1mm}
-c_x+2a-2Qb=0.
\endmatrix 
\right.
$\hss
\hbox to 0pt {\hss
$\displaystyle \matrix (2.3a)\\
\vspace {1mm}
(2.3b)\\
\vspace {1mm}
(2.3c)\endmatrix $}}
\vskip2mm
\noindent
The equalities (2.3b) and (2.3c) demand equivalently 
$$\left \{\aligned & a=\frac14 (b+c)_x+\frac12 Q(b+c),\\
\vspace {1mm}
&(b-c)_x-2Q(b-c)=0.\endaligned\right.$$
From the second equality above, 
we obtain that $b-c=[b(x_0)-c(x_0)]\text{exp}
\bigl(\int_{x_0}^{x}2Q\,dx\bigr)$.
We are interested in the most simple case $b=c$. For this case, we
have $a=\frac12 c_x +Qc$ and (2.3a) reads as
$$Q_t=a_x=\part (\frac12 \part +Q)c.\tag2.4$$

Assume that 
$$V=V^{(m)}=
(\frac 12 c^{(m)}_x+Qc^{(m)})
\si _1 +
c^{(m)}i\si _2 +
c^{(m)}\si _3, \ c^{(m)}
=\sum _{j=0}^m c_j\la ^{m-j},\ m\ge
0.\tag2.5$$
Further for the sake of convenience, set 
$$R_0=\frac12 \part +v_0,\ R_i=v_i,\ 1\le i\le q. \tag2.6$$
In this way, we have
$$\align &\part (\frac12 \part +Q)c^{(m)}=\part \sum _{i=0}^mR_i\la
^i \sum _{j=0}^mc_j\la ^{m-j}\\
=& \part \sum
_{i=0}^{q-1}(
\sum _{j=0}^i
R_j c_{j+m-i})\la ^{i}
+\part
 \sum_{i=q}^{q+m}(
\sum _{j=0}^q
R_j c_{j+m-i})\la ^{i}
,\tag2.7\endalign$$
where $c_j=0,\ j<0$. 
In order to generate isospectral integrable systems from (2.4), 
it is required that 
$$\part 
\sum _{j=0}^q
R_j c_{j+m-i}=0,\ q\le i\le  q+m.$$
These equations have a special solution
$$
\sum _{j=0}^q
R_j c_{j+m-i}=\cases -1, &i=q+m,\\
0,& q\le i\le q+m-1.\endcases
$$
Noticing that $R_q=-1$, this relation 
can be satisfied by choosing
$$c_0=1,\ c_j=
\sum_{i=0}^{q-1} R_ic_{i+j-q},\ 
1\le j\le \infty.\tag2.8$$
It follows that
$$ c_1=v_{q-1},\ c_2=v_{q-2}+v_{q-1}^2.$$
A general solution $\{c_j\}$
is a linear combination of the solution (2.8) with arbitrary functions
of $t$ as coefficients. It couldn't lead to new integrable systems
in nature.

At this stage,
we see from (2.7) that (2.4) engenders equivalently the isospectral 
 $(\la_{t_m}=0)$  integrable hierarchy 
$$\align &u_{t_m}=K_m=\part (R_0c_m,R_0c_{m-1}+R_1c_m,\cdots, 
\sum _{j=0}^{q-1} R_jc_{j+m-(q-1)})^T
\\&=\part (R_0c_m,R_0c_{m-1}+R_1c_m,\cdots, 
\sum _{j=0}^{q-2} R_jc_{j+m-(q-2)},c_{m+1})^T
,\ m\ge0,\tag2.9\endalign$$
and the isospectral hierarchy possesses the Lax pairs
$$\align \phi_x= &U\phi=
 (Q\si _1+i\si _2+\si _3)\phi=\left (\matrix 1
&Q+1\\ Q-1& -1 \endmatrix \right )\phi,\tag2.10a\\ 
\phi_{t_m}=&V^{(m)}\phi=
\bigl ((
\frac 12 c^{(m)}_x+Qc^{(m)}
) \si _1 +
c^{(m)}i\si _2 +
c^{(m)}\si _3\bigr)\phi\\ \vspace {1mm}
=&
\pmatrix 
 c^{(m)}& 
\frac 12 c^{(m)}_x+c^{(m)}(Q+1)\\ \vspace {1mm}
\frac 12 c^{(m)}_x+c^{(m)}(Q-1)
&- c^{(m)}\endpmatrix\phi,\tag2.10b\endalign$$
where $ c^{(m)} =\sum _{j=0}^m c_j\la ^{m-j}$, with the $c_j$ defined
by (2.8). 
To show that the isospectral hierarchy (2.9) exhibits a kind of 
hereditary structure, we define an integro-differential operator
$$ \Phi=
\left (\matrix 0 & 0 &\cdots & 0 & P _0 \\
 1 & 0 &\cdots & 0 & P _1 \\
 0 & 1 &\cdots & 0 & P _2 \\ \vdots &\vdots & \ddots &\vdots
&\vdots \\
0 & 0 &\cdots & 1 & P _{q-1} \endmatrix \right ) ,
\ \ 
P _0=\frac12\part +\part v_0\part ^{-1},\ P_i=\part v_i\part ^{-1},\ 
1\le i\le q.
\tag2.11$$
Since, as is easily seen, 
$P_q=-1,\ P_i\part =\part R_i,\ 0\le i\le q,$  we
obtain
$$ u_{t_m} =K_m=\Phi K_{m-1}=\cdots =\Phi ^m K_0=\Phi ^m u_x,\
m\ge0,\tag2.12$$
that is to say,
$$\left (\matrix v_{0t_m}\\v_{1t_m}\\ v_{2t_m}\\
\vdots\\v_{q-1,t_m}\endmatrix \right )
=
\left (\matrix 0 & 0 &\cdots & 0 & \frac12 \part +\part v_0\part ^{-1} \\
 1 & 0 &\cdots & 0 & \part v_1\part ^{-1} \\
 0 & 1 &\cdots & 0 & \part v_2\part ^{-1} \\ \vdots &\vdots & \ddots &\vdots
&\vdots \\
0 & 0 &\cdots & 1 & \part v _{q-1}\part ^{-1} \endmatrix \right )
^m
\left (\matrix v_{0x}\\v_{1x}\\ v_{2x}\\
\vdots\\v_{q-1,x}\endmatrix \right )
,\ m\ge0.\tag2.13$$
Noticing the special form $\part v_i \part ^{-1}$ in $P_i$,
we see from the nice form (2.13) 
itself that every system in the hierarchy (2.13)
is {\it local} in spite of the integro-differential character of the
operator $\Phi $. 
Multi-component integrable systems in (2.13) 
are similar, in appearance, to coupled KdV
systems in 7). We shall to show in Section 4 that 
the flows generated by (2.13) commute with each other.
 But we shall also see later
that these flows have different characteristics from those of coupled KdV
systems.

\vskip 4mm
\centerline{{\bf 3. Nonisospectral integrable systems}} 
\vskip 3mm

Let us now turn to  the deduction of nonisospectral ($\la _{t_n}=\la
^{n+1}$) integrable systems associated with the spectral problem (2.1).
In order  to apply the generating scheme in Ref. 8), we should verify 
that the characteristic operator equation 
$$ [\Omega ,U]+\Omega _x=U'[\Phi X]-\la U'[X]\tag3.1$$
has solutions $\Omega=\Omega(X)$
for any given vector field $X=(X_1,\cdots,X_q)^T$
and need to search for a pair of solutions 
$B_0,\,g_0$ to certain key operator equation
$$[B_{0},U]+B_{0x}=
U'[g_0]+\la ^kU_\la=
U'[g_0]+\la U_\la \ (k=1) .\tag3.2$$
Here the key operator equation with $k=1$ is imposed.
In view of the former deduction of isospectral integrable systems, we may
suppose that $\Omega,\, B_0$ possess the forms
$$\align  \Omega =&
\bigl (( \frac 12 c(\Omega)_x +Q c(\Omega) ) \si _1 +
c(\Omega) i\si _2 + c(\Omega) \si _3\bigr)\\=&
\pmatrix  c(\Omega) & \frac 12
c(\Omega)_x + c(\Omega)(Q+1) \\ 
\vspace {1mm}
\frac 12  c(\Omega)_x
+ c(\Omega)(Q -1) &-  c(\Omega) \endpmatrix,
\\  B_0=&\bigl (( \frac 12 c(B_0)_x + Qc(B_0) ) \si _1 +
c(B_0) i\si _2 + c(B_0) \si _3\bigr)\\=&
\pmatrix  c(B_0) & \frac 12
c(B_0)_x + c(B_0)(Q +1) \\ 
\vspace {1mm}
\frac 12  c(B_0)_x
+ c(B_0)(Q -1) &-  c(B_0) \endpmatrix,
\endalign$$
where $c(\Omega),\,c(B_0)$ are two functions to be determined.
Set $\Phi X=\bigl((\Phi X)_1,\cdots,(\Phi X)_q\bigr)^T,$ 
$g_0=(g_{01}, \cdots,$ $g_{0q})^T.$
Then we have 
$$\align &U'[\Phi X]-\la U'[X]=\sum _{i=1}^q \bigl[ (\Phi X)_i-\la
X_i\bigr]\la ^{i-1}\si _1=\bigl[\sum _{i=0}^{q-1}(P_i X_q)\la
^i-X_q\la ^q\bigr]\si _1,\\ &
U'[g_0]+\la U_\la =\big [\sum _{i=0}^{q-1}(g_{0i}+iv_i)\la ^i\bigr]\si
_1.\endalign$$
Now we easily find that $c(\Omega )=\frac12 \part ^{-1}X_q,\, 
c(B_0)=qx$. Therefore we obtain the following solution to (3.1)
$$\align \Omega =&
\bigl (( \frac 14 X_q + \frac12 Q\part ^{-1}X_q ) \si _1 +
\frac12 \part ^{-1}X_q  i\si _2 + \frac12 \part ^{-1}X_q  \si _3\bigr)\\
\vspace {1mm}
=& \pmatrix  \frac12 \part ^{-1}X_q  & \frac 14
 X_q  + \frac12(Q+1) \part ^{-1}X_q 
  \\ 
\vspace {2mm}
\frac 14 X_q 
+ \frac12 (Q-1)\part ^{-1}X_q  & -  \frac12 \part ^{-1}X_q  \endpmatrix,
\tag3.3
\endalign $$
and the following pair of solutions to (3.2)

\vskip 2mm
\line 
{\hbox to 0pt {\hss}
\hss
$\displaystyle\left \{
\matrix \format \l \\
B_0=\bigl (( \frac 12 q +Q qx ) \si _1 +
qx i\si _2 + qx \si _3\bigr)=
\pmatrix  qx & \frac 12
q +qx(Q  + 1) \\ 
\vspace {2mm}
\frac 12 q
+qx(Q  -1) &-  qx \endpmatrix,\\
\vspace {1mm}
g_0=\bigl( qv_0,(q-1)v_1,\cdots, v_{q-1}\bigr)^T+qx 
\bigl( v_0,v_1,\cdots, v_{q-1}\bigr)^T_x . 
\endmatrix\right.
$\hss
\hbox to 0pt {\hss 
$\displaystyle
\matrix (3.4a)\\
\vspace {3mm}
(3.4b)\endmatrix$
}}
\vskip 2mm
\noindent Now taking full advantage of
 the result in Ref. 8), we know that
the Lax pairs 

\vskip 2mm
\line 
{\hbox to 0pt {\hss}
\hss
$\displaystyle\left \{
\matrix \format \l \\
\phi_x=U\phi,\ \la _{t_n}=\la ^{n+1},\\ \vspace {2mm}
\phi_{t_n}=W^{(n)}\phi,\ W^{(n)}=\la ^nB_0+\sum _{j=1}^n \la
^{n-j}\Omega (\Phi^{j-1}g _0),
\endmatrix\right.
$\hss
\hbox to 0pt {\hss 
$\displaystyle
\matrix  (3.5a)\\
\vspace {4mm}
(3.5b)\endmatrix$
}}
\vskip 2mm
\noindent lead to the following nonisospectral ($\la _{t_n}=\la
^{n+1})$ hierarchy of multi-component integrable systems
$$u_{t_n}=\rho _n=\Phi ^n g _0,\ n\ge 0,\tag3.6$$
that is to say, $$
\left (\matrix v_{0t_n}\\v_{1t_n}\\ v_{2t_n}\\
\vdots\\v_{q-1,t_n}\endmatrix \right )
=
\left (\matrix 0 & 0 &\cdots & 0 & \frac12 \part +\part v_0\part ^{-1} \\
 1 & 0 &\cdots & 0 & \part v_1\part ^{-1} \\
 0 & 1 &\cdots & 0 & \part v_2\part ^{-1} \\ \vdots &\vdots & \ddots &\vdots
&\vdots \\
0 & 0 &\cdots & 1 & \part v _{q-1}\part ^{-1} \endmatrix \right )
^n
\left [\left (\matrix qv_{0}\\(q-1)v_{1}\\ (q-2)v_{2}\\
\vdots\\v_{q-1}\endmatrix \right )+
qx\left (\matrix v_{0x}\\ v_{1x}\\ v_{2x}\\
\vdots\\ v_{q-1,x}\endmatrix \right )\right ]
,\ n\ge0.\tag3.7$$
These nonisospectral systems are nonlocal except the first one.
In the next section, we shall show that (3.7) yields  a hierarchy of 
common first order master-symmetries 
and thus Virasoro symmetry algebras are generated 
for the isospectral integrable systems 
(2.13).

\vskip 4mm
\centerline {{\bf 4. Virasoro symmetry algebras}}
\vskip 3mm

In what follows, we would like  to give Virasoro
 $\tau $-symmetry algebras for
the isospectral hierarchy (2.13).
Towards this end, let us first prove that $\Phi $ in (2.11) is a
hereditary symmetry$^{9)}$ (or Nijenhuis operator$^{10)}$).
Namely, we must verify that 
$\Phi '[\Phi K]S-\Phi\Phi '[K]S$ is symmetric with respect to
two vector fields $K,\,S$.
In fact, for $K=(K_1,\cdots,K_q)^T,\,S=(S_1,\cdots,S_q)^T$,
we can compute that 
$$\align &
\Phi '[\Phi K]S-\Phi\Phi '[K]S\\ =&\underset i+1 \to {(\cdots, 
\part (P_iK_q)\part
^{-1}S_q-P_i\part K_q\part ^{-1} S_q,\cdots)^T},\ 0\le i\le q-1,\\
& 
\part (P_iK_q)\part
^{-1}S_q-P_i\part K_q\part ^{-1} S_q\\
=&-\frac12 \delta _{i0}(K_{qx}S_q+K_qS_{qx})+v_{ixx}(\part
^{-1}K_q)(\part ^{-1}S_q)+v_{ix}\bigl[K_q(\part ^{-1}S_q)+(\part
^{-1}K_q)S_q\bigr],\endalign$$
whereupon $\Phi$ is a hereditary symmetry, indeed. 
Secondly, a direct computation can show that
$$[K_0,g _0]=qK_0,\ L_{K_0}\Phi=0,\ L_{g _0}\Phi =\Phi,\tag4.1$$
where the commutator of vector fields is defined by
$[K,S]=K'[S]-S'[K]$
and the Lie derivative, by $L_K\Phi=\Phi'[K]-[K',\Phi]$ (see
Ref. 11) for more information).
Based on the basic result in Ref. 11), 
these two properties permit us to conclude  that
the $s$th isospectral coupled  
integrable system $u_{t_m}=\Phi ^su_x$ has 
two hierarchies of $K$-symmetries (time independent symmetries)
and $\tau$-symmetries (time first order dependent  symmetries)
$$K_m=\Phi ^m u_x,\ m\ge0;\ \tau_n^{(s)}=
(s+q)tK_{n+s}+\rho _n,\ n\ge0,\tag4.2$$
and that these symmetries constitute
a $\tau$-symmetry algebra
\vskip 2mm
\line
{\hbox to 0pt {\hss}
\hss $\displaystyle
\left \{\matrix \format \l\\
[K_m,K_n]=0,\ m,n\ge0,\\
\vspace {1mm}
[K_m,\tau ^{(s)} _n]=(m+q)K_{m+n},\ m,n\ge 0,\\
\vspace {1mm}
[\tau _m^{(s)},\tau_n^{(s)}]=(m-n)\tau _{m+n}^{(s)},\ m,n\ge0.
\endmatrix 
\right.
$\hss
\hbox to 0pt {\hss
$\displaystyle \matrix (4.3a)\\
\vspace {2mm}
(4.3b)\\ 
\vspace {2mm}
(4.3c)\endmatrix $}}
\vskip2mm
\noindent
This symmetry
algebra also shows that $\rho _n,\,n\ge0,$ defined by (3.6), 
are the common first order
master-symmetries$^{12)}$ of the integrable hierarchy (2.13).

\vskip 4mm
\centerline{{\bf 5. Conclusions and Remarks}}
\vskip 3mm

We have constructed a hierarchy of coupled  integrable systems
(2.13) and each system in the hierarchy (2.13) has a $\tau$-symmetry
algebra (4.3), in which master symmetries are generated by nonisospectral 
flows of the spectral problem (2.1).
The first nonlinear system in the integrable hierarchy (2.13)
reads as 
$$\left \{
\aligned &
v_{0t_1}=\frac12 v_{q-1,xx}+(v_0v_{q-1})_x,\\&
v_{1t_1}=v_{0,x}+(v_1v_{q-1})_x,\\&\quad \hdots\hdots \\&
v_{q-2,t_1}=v_{q-3,x}+(v_{q-2}v_{q-1})_x,\\&
v_{q-1,t_1}= v_{q-2,x}+(v_{q-1}^2)_x.\endaligned \right.\tag5.1
$$
This resembles coupled Burgers equations but
it seems to us that 
it couldn't be simplified to Burgers equations. Also, it might be 
very difficult
to find a reduction of the integrable hierarchy (2.13) 
to scalar Burgers equations $w_{t_m}=(\frac 12 \part +\part w\part
^{-1})^mw_x,\ m\ge0.$ 

When $q=2$, the resulting isospectral hierarchy (2.13) becomes
$$\pmatrix v_0\\
\vspace {1mm}
v_1\endpmatrix _{t_m}=\Phi^m
\pmatrix v_0\\
\vspace {1mm}
v_1\endpmatrix _{x},\ \Phi=\pmatrix 0&\frac12 \part
+\part v_0\part ^{-1}\\
\vspace {1mm}
1&\part v_1\part ^{-1}\endpmatrix,\ m\ge0.\tag5.2$$
The first two nonlinear systems
of the hierarchy (5.2) are as follows
$$\left \{\aligned 
&v_{0t_1}=\frac12 v_{1xx}+(v_0v_1)_x,\\&
v_{1t_1}= v_{0x}+2v_1v_{1x};\endaligned\right.\tag5.3$$
$$\left \{\aligned 
&v_{0t_2}=\frac12 v_{0xx}+(v_1v_{1x}+v_0^2+v_0v_1^2)_x,\\&
v_{1t_2}= \frac12 v_{1xx}+(2v_0v_{1}+v_1^3)_x.\endaligned\right.\tag5.4$$
The first system has already appeared in Ref. 6) and 
can also be solved by means of WTC
Painlev\'e analysis method$^{13)}$.
Moreover, the system (5.3) can be expressed as 
$$\left ( \matrix v_{0t_1}\\v_{1t_1}\endmatrix \right)
=J_1G_1=
\left (\matrix \part & 0\\0&\part \endmatrix \right)
\left ( \matrix \frac12 v_{1x}+v_0v_1\\v_0+v_{1}^2\endmatrix \right)
=J_2G_2=
\left ( \matrix 0&\part\\ \part&0 \endmatrix \right)
\left ( \matrix 
v_0+v_{1}^2\\
\frac12 v_{1x}+v_0v_1
\endmatrix \right).\tag5.5$$
Although $J_1,\,J_2$ are all Hamiltonian operators, $G_1,\,G_2$ aren't 
gradients, i.e.
$$G_i\ne \frac {\delta H_i}{\delta u},
\ H_i=\int _0^1 u^TG_i(\la u)\,
d\la ,\ u=(v_0,v_1)^T,\ i=1,2.$$
Therefore the system (5.3) doesn't possess Hamiltonian
structures with the form (5.5). 
With the same argument, it may be verified that the system
(5.4) doesn't possess similar Hamiltonian structures, either.

In general, the isospectral hierarchy (2.13)
doesn't possess Hamiltonian structures with a simple Hamiltonian 
operator as above. This property is completely 
different from those of usual systems of soliton equations, for example, 
coupled KdV systems$^{7)}$ and WKI systems$^{14)}$, both of which 
possess Hamiltonian structures. Moreover,
we conjecture that there exists
a finite number of polynomial conserved densities for
the hierarchy (2.13) and that each system in the hierarchy (2.13) is
$C$-integrable (for definition, see Ref. $15)\,$). It is
also interesting for us
to check whether or not the hierarchy (2.13) may be derived from 
Sato theory$^{3)}$ as a reduction or may be transformed
into a bilinear form$^{16)}$.

\vskip 0.5cm
\noindent {\bf 
Acknowledgments:} 
This work was supported by the 
Alexander von Humboldt Foundation of Germany, the National Natural Science
Foundation of China, the 
Shanghai Science and Technology Commission of China and Fok Ying-Tung 
Education Foundation of China.
One (W. X. Ma) of the authors would like to thank Prof. 
B. Fuchssteiner and Dr. W. Oevel for valuable discussions at Soliton 
Seminar in University of Paderborn, Germany.

\newpage
\vskip 5mm
\centerline {{\bf References}}
\vskip 3mm

\item{1)} P. D. Lax, Commun. Pure Appl. Math.
{\bf 21} (1968), 467.
\item{2)} F. Calogero and A. Degasperis, Lett. Nuovo Cimento
{\bf 22} (1978), 131, 263, 420;
W. L. Chan and Y. K. Zheng, Lett. Math. Phys. {\bf 14}
(1987), 293; W. L. Chan and K. S. Li, J. Math. Phys. 30 (1989) 2521. 
\item{3)} E. Date, M. Kashiwara, M. Jimbo and T. Miwa, in {\it 
Nonlinear Integrable Systems-Classical and Quantum Theory}, ed. M. Jimbo and 
T. Miwa (World Scientific, Singapore, 1983), p. 39;
 M. Jimbo and T. Miwa, Publ. RIMS, Kyoto Univ. {\bf 19} (1983), 943;
Y. Ohta, J. Satsuma, D. Takahashi and T. Tokihiro,
Prog. Theor. Phys. Suppl. No. 94 (1988), 210.
\item{4)} M. Toda, Physica D {\bf 33} (1988), 317.
\item {5)} C. H. Gu and H. S. Hu, Lett. Math. Phys.
{\bf 11} (1986), 325;
Z. X. Zhou, Phys. Lett. A {\bf 168} (1992), 370.
\item {6)} X. G. Guo,  Phys. Lett. A {\bf 147} (1990), 491.
\item {7)} M. Boiti, P. J. Caudrey and F. Pempinelli, 
Nuovo Cimento B {\bf 83} (1984), 71; M. Antonowicz and A. P. Fordy,
Physica D {\bf 28} (1987), 345; W. X. Ma, J. Phys. A: Math. Gen. {\bf 26}
(1993), 2573.
\item {8)} W. X. Ma, Phys. Lett. A {\bf 179} (1993), 179;
 J. Phys. A: Math. Gen. {\bf 25} (1992), L719. 
\item {9)} B. Fuchssteiner,
Nonlinear Anal. Theor. Meth. Appl. {\bf 3} (1979), 849.
\item {10)}
F. Magri, in {\it Nonlinear Evolution Equations and
Dynamical Systems}, Lecture Notes in Physics Vol. 120, eds. M. Boiti,
F. Pempinelli  and  G. Soliani
(Springer-Verlag, Berlin, 1980), p. 233.
\item {11)} 
W. Oevel,  in {\it Topics in Soliton Theory and Exactly
Solvable Nonlinear Equations}, eds. M. J. Ablowitz, B. Fuchssteiner and
M. D. Kruskal (World Scientific, Singapore, 1987), p. 108;
A. S. Fokas and P. M. Santini,  in: {\it Symmetries and
Nonlinear Phenomena} eds. D. Levi and P. Winternitz 
(World Scientific, Singapore, 1988), p. 7;
W. X. Ma, J. Phys. A: Math. Gen. {\bf 23} (1990), 2707.
\item {12)} B. Fuchssteiner,  Prog. Theor. Phys.
{\bf 70} (1983), 1508.
\item {13)} J. Weiss, M. Tabor and G. Carnevale, 
J. Math. Phys. {\bf 24} (1983), 522.
\item{14)} M. Wadati, K. Konno and Y. H. Ichikawa, J. Phys. Soc. Jpn. 
{\bf 47} (1979), 1698.
\item {15)} F. Calogero and W. Eckhaus, 
 Inverse Problems {\bf 3} (1987), 229.
\item{16)} R. Hirota, in {\it Solitons}, Topics in Current Physics 17, 
eds. R. K. Bullough and P. J. Caudrey
(Springer-Verlag, Berlin, 1980), p. 157.

\bye